\documentclass[12pt]{article}
\usepackage[top=2cm,left=2cm,right=2cm]{geometry}
\usepackage[inline]{showlabels}

\usepackage[numbers,square]{natbib}

\usepackage{axodraw}
\usepackage{epsfig}
\usepackage{cancel}
\usepackage{caption}
\usepackage{feynmf} 
\usepackage{amssymb}
\usepackage{amsfonts}
\usepackage{epsf}
\usepackage{rotating}
\usepackage{graphicx}
\usepackage{amsmath}
\usepackage{fancyhdr}
\usepackage{subfigure}
\usepackage{graphics}
\usepackage{pstricks}
\usepackage{color}
\usepackage{frontespizio}
\usepackage{hyperref}
\hypersetup{
    colorlinks,
    citecolor=green,
    filecolor=black,
    linkcolor=blue,
    urlcolor=black
}
\usepackage{type1ec}
\usepackage[T1]{fontenc}
\usepackage{lettrine}
\usepackage{bbold}
\usepackage{calligra}
\usepackage{tikz}
\usepackage{subfigure}
\usepackage{mathrsfs}
\usepackage{curve2e}
\usepackage{setspace}
\usepackage{indentfirst}
\usepackage{emptypage}
\usepackage[babel]{csquotes} 
\usepackage[font=small,labelfont=bf,labelsep=quad]{caption} 
\usepackage{graphicx} 
\usepackage{listings} 
\usetikzlibrary{patterns}
\usepackage{relsize}
\usetikzlibrary{intersections,positioning}
\usetikzlibrary{decorations.pathmorphing,decorations.markings,arrows,positioning}
\usepackage{braket}
\usepackage{mathrsfs}
\usepackage{stackengine}
\usepackage{calc}
\newlength\shlength
\newcommand\xshlongvec[2][0]{\setlength\shlength{#1pt}%
  \stackengine{-5.6pt}{$#2$}{\smash{$\kern\shlength%
    \stackengine{7.55pt}{$\mathchar"017E$}%
      {\rule{\widthof{$#2$}}{.57pt}\kern.4pt}{O}{r}{F}{F}{L}\kern-\shlength$}}%
      {O}{c}{F}{T}{S}}

\newcommand{\sdfrac}[2]{\mbox{\small$\displaystyle\frac{#1}{#2}$}}

\IfFileExists{dsfont.sty}
{\usepackage{dsfont}
	\let\mathbb=\mathds
	\newcommand{\id}{\mathds{1}}}
{\typeout{Package dsfont.sty was not found, using alternative macros.}
	\let\mathds=\mathbb
	\newcommand{\id}{\mbox{1 \kern-.59em {\rm l}}}}

\usepackage{slashed}
\usepackage{units}
\usepackage{setspace}
\newcommand{\eeqa}{\end{eqnarray}}
\newcommand{\beqa}{\begin{eqnarray}}

\newcommand{\nn}{\nonumber}

        \let\m=\mu
\let\n=\nu

\let\D=\Delta

\def\nbox#1#2{\vcenter{\hrule \hbox{\vrule height#2in
			\kern#1in \vrule} \hrule}}
\def\sq{\,\raise.5pt\hbox{$\nbox{.09}{.09}$}\,}
\def\sqb{\,\raise.5pt\hbox{$\overline{\nbox{.09}{.09}}$}\,}

\newcommand{\bea}{\begin{eqnarray}}
\newcommand{\eea}{\end{eqnarray}}
\newcommand{\be}{\begin{equation}}
\newcommand{\ee}{\end{equation}}

\newcommand{\bes}{\begin{subequations}}
	\newcommand{\ees}{\end{subequations}}

\def\nn{\nonumber\\}

\numberwithin{equation}{section}

\usepackage{accents}

\begin{document}
\begin{center}
\vspace{1.5cm}
{\Large\bfseries
Renormalization, Conformal Ward Identities\\ and the Origin  of a Conformal Anomaly Pole\\ }
\vspace{0.3cm}
{\Large\bfseries  }
\vspace{0.2cm}

\vspace{0.3cm}
{\Large\bfseries}

\vspace{0.3cm}

\vspace{2 cm}

{\bf  Claudio Corian\`o and Matteo Maria Maglio \\}

\vspace{0.5cm}

{\em Dipartimento di Matematica e Fisica "Ennio De Giorgi"\\ Universit\`a del Salento and INFN Lecce, \\
Via Arnesano, 73100 Lecce, Italy}

\vspace{0.5cm}

\end{center}

\begin{abstract}
We investigate the emergence of a conformal anomaly pole in conformal field theories in the case of the $TJJ$ correlator. We show how it comes to be generated in dimensional renormalization, using a basis of 13 form factors (the $F$-basis), where only one of them requires renormalization $(F_{13})$, extending previous studies. We then combine recent results on the structure of the non-perturbative solutions of the conformal Ward identities (CWI's) for the $TJJ$ in momentum space, expressed in terms of a minimal set of 4 form factors ($A-$ basis), with the properties of the $F$-basis, and show how the singular behaviour of the corresponding form factors in both basis can be related. The result proves the centrality of such massless effective interactions induced by the anomaly, which have recently found realization in solid state, in the theory of topological insulators and of Weyl semimetals. This pattern is confirmed in massless abelian and nonabelian theories (QED and QCD) investigated at one-loop.

\newpage
\end{abstract}
\section{Introduction}
Chiral and conformal anomalies are central in quantum field theory, due to the appearance in anomaly vertices of non-conserved chiral or dilatation currents. 
Conditions of gauge anomaly cancellations - for gauge anomalies - and/or the identification of possible global anomalies, play a key role in determining the particle spectra of the corresponding theories, constraining their quantum numbers.\\
 In general, most of the analysis has always been associated with the investigation of the Ward identities (WI) of a given anomalous correlator, in the form of conservation  - for chiral - or trace and conservation WI's for conformal anomalies. These operations reduce the number of free uncontracted  
indices of an anomalous diagram and mix their defining tensor components and form factors, providing less information with respect to that which is obtainable from the study of a full (uncontracted)  vertex.\\
It has been shown that in an uncontracted anomaly vertex of either chiral, conformal \cite{Giannotti:2008cv,2009PhLB..682..322A, Armillis:2009pq, Armillis:2010qk} or superconformal type \cite{Coriano:2014gja}, the origin of an anomaly has to be attributed  to the appearance of specific form factors in its tensor structure, which are 
proportional to $1/k^2$ in the massless limit. Such anomaly poles define massless exchanges in momentum space and are the direct signature of the anomaly. In all such cases $k$ denotes the momentum of an axial-vector current in an $AVV$ (axial-vector/vector/vector) correlator or that of a stress energy tensor ($T$) in a TJJ vertex.\\
Previous studies in perturbation theory, away from the conformal limit, by the inclusion, for instance, of a fermion mass in the loop, have shown that the form factors which appear in the trace part of the $TJJ$ correlator are characterized by spectral densities which satisfy mass-independent conformal \cite{Giannotti:2008cv} and, in the supersymmetric case, superconformal \cite{Coriano:2014gja} sum rules, related to the anomaly coefficients. In the massless  fermion limit their spectral densities converge to $\delta$-functions, manifesting the exchange of an anomaly pole. This beautiful behaviour, obviously, is not just a coincidence and suggests of 
something very special taking place in the conformal/chiral anomaly actions.\\
The existence of chiral anomaly poles has been discussed in the 
literature since the work of Dolgov and Zakharov \cite{Dolgov:1971ri}, while conformal anomaly poles have been shown to be part of the $TJJ$  vertex in QED \cite{Giannotti:2008cv, Armillis:2010qk}, QCD and the electroweak sector of the Standard Model \cite{Coriano:2011ti, Coriano:2011zk} only more recently. In the case of the Standard Model it has been argued that an effective dilaton-like interaction could be mediated by the trace anomaly, due to such massless exchanges, which could be of phenomenological interest at the LHC \cite{2013JHEP...06..077C, Bandyopadhyay:2016fad}.  It is then natural to interpret such intermediate states 
as the signature of (anomaly) broken scale invariance of the Higgs sector, if the zero mass limit of the Higgs sector is taken \cite{2013JHEP...06..077C}. It is an open question, in the supersymmetric context, for instance, 
if the three anomaly poles of the superconformal currents supermultiplet, which interpolate with an axion-dilaton-dilatino (ADD) composite multiplet, are an indication of the possible existence of a broken conformal phase in $\mathcal{N}=1$ supersymmetric theories in which supersymmetry is nonlinearly realized  \cite{Coriano:2014gja}.

Studies of such interactions in the context of both chiral and conformal anomaly diagrams have always been performed at the perturbative level in the past, with the obvious limitations of the case. These studies 
show the presence of some universal features of these interactions, confirming that anomaly poles are ubiquitous in the presence of anomalous interactions. The chiral and conformal anomaly coefficients are then proportional to the residues of the corresponding correlators evaluated at the anomaly pole (times a tensor structure which is the anomaly functional). \\
  In the case of a global $U(1)_B$ anomaly, with an external $B_\mu$ gauge field and field strength $F_{B \, \mu\nu}$, coupled to anomalous (axial-vector, $A$)  current, the anomaly action can indeed be written in the generic form 
 \be
a_n\int d^4 x\, d^4 y\partial\cdot B (x) \left(\frac{1}{\square}\right)(x,y)F_B\tilde{F}_B(y)
  \ee
 for a chiral anomaly  of ${U(1)}_B^3$  type, corresponding to a vertex with three axial vector currents ($AAA$) and anomaly coefficient $a_n$. A similar behaviour is expected from a gravitational anomaly $(a_g)$, generated at an axial-vector/graviton/graviton (ATT) vertex 
 \be
a_g\int d^4 x\, d^4 y\partial\cdot  B(x)  \left(\frac{1}{\square}\right)(x,y) R\tilde{R}(y)
  \ee
with $R_{\mu\nu\alpha\beta}$ being the Riemann tensor, when coupling an axial-vector current mediated by an external pseudoscalar gauge field $B_\mu$ to two stress-energy tensors. A third example is provided by the $TJJ$ vertex, that we are going to discuss below, which is affected by a conformal anomaly and manifests a similar interaction. \\
Understanding the key role played by these effective interactions at any energy scale, and in the presence of radiative effects which may corrupt their massless behaviour, is crucial for a more complete comprehension of their dynamics. In particular, their emergence in the anomaly effective action calls for a more physical re-interpretation of the irreversibility of RG-flows from the UV to the IR, in theories with conformal anomalies, which should be described, on physical grounds, also in terms of such effective interactions. The appearance of the $\beta$ function at the numerator of an anomaly pole, and its dependence on the number of massless degrees of freedom along the flow, is clearly an indication that such possibility should not be excluded.

\subsection{Poles in special and general kinematics}  
There are reasons why these contributions have been overlooked in the past, and they have to do with the proliferation of tensor structures of such vertices, as exemplified in the case of the $AVV$ diagram, for which at least two most valuable representations exist. Most notably, these are the Rosenberg representation \cite{Rosenberg:1962pp}, which is expressed in terms of 6 form factors, that reduce to 4 by applications of the vector Ward identities, and the longitudinal/transverse (L/T) parameterization \cite{Knecht:2002hr}, used in the analysis of the anomalous magnetic moment of the muon. In the latter case a complete 2-loop computation has shown the non-renormalizability of the entire vertex \cite{Jegerlehner:2005fs}\cite{Vainshtein:2002nv}, not just of its longitudinal part, as one would expect from the Adler-Bardeen theorem \cite{Adler:1969er}. \\ 
The issue of whether poles are genuine or artificially introduced by a certain {\em ad hoc} parameterization of a given vertex has generated wide disagreement over the years, and it has also been a source of confusion. In fact, in general, an anomalous correlator has extra poles beside the anomaly poles. Therefore in order to make a distinction between an anomaly pole and the remaining (non anomalous) poles present in its (several) tensor structures, requires an in-depth study of the corresponding Feynman diagram. 
If some parameterizations obscure the pole behaviour, as in Rosenberg's formulation, in others, such as the L/T one, the pole is present for any momenta of the vertex. \\
One possible way to resolve such a dispute is to go beyond perturbation theory, if possible, using exact results 
if these are available. Such is the case of conformal field theories (CFT's) where the presence of extra conformal Ward identities (CWI's) - with respect to Poincare' invariance - allow to specify, at least for some correlators, their momentum dependence.  

The goal of the present work is to illustrate how a pole emerges from the renormalization of a single form factor $(F_{13})$ in a specific (non minimal) basis of the $TJJ$ vertex. The result holds in general for any $TJJ$ vertex in CFT. We show how to combine such information with a recent analysis of the solutions of CWI's based on a (minimal) basis of form factors $(A_1,\ldots A_4)$ fixed by conformal symmetry. Our results rely on recent solutions of the conformal equations presented in \cite{Bzowski:2013sza, Bzowski:2017poo} and prove that the emergence of a specific pole in the correlator is not the result of redundancy or an artefact of the parameterization. It should rather be thought of as a conclusive manifestation of an anomaly, and it is not strictly associated to a specific configuration of the external momenta of a vertex, but holds also off-shell.  \\
In this work we will concentrate on the physical implications of our analysis, leaving aside the 
rather technical parts that will be presented in a companion paper and in work in preparation. We are going to show how the combination of perturbative results in massless QED and of non-perturbative information derived from the solution of the  conformal Ward identities, allows us to trace back how 
an anomaly pole appears in a simpler correlator such as the $TJJ$. \\ 
Our conclusions will be that the anomaly pole of the $TJJ$ is a crucial part of this anomalous interaction. We believe that similar conclusions can be drawn in all cases in the corresponding anomaly actions. 

\subsection{The perturbative analysis} 
In momentum space, the emergence of these poles can be attributed to a specific configuration of the loop integral in the Feynman expansion of the correlator, with the exchange of a (fermion/ gluon) collinear pair 
\cite{Giannotti:2008cv,Armillis:2009pq,Armillis:2010qk}.
 Anomaly poles trigger virtual interactions which redefine the vacuum of a theory and, in a simple perturbative analysis, cannot be immediately recognized as asymptotic states of an effective S-matrix. Rather they can be thought of as effective intermediate exchanges mediated by an anomaly. The solution of the conformal constraints that we are going to present in this work indicates that such viewpoint and limitations are a consequence of perturbation theory. In this respect, the nonperturbative approach provided by the solutions of the CWI's shows that such pole-like contributions are generically present in the off-shell anomaly vertex. \\
 This suggests that theories affected by anomaly poles may undergo a non-perturbative redefinition of their vacuum in such a way that such interactions may describe, in a non-perturbative phase of such theories,  the exchange of composite (asymptotic) states in the infrared, with specific quantum numbers. 
 This transition requires a mechanism of dynamical breaking of a conformal symmetry, of which the anomaly is probably just one component.   \\
 The most interesting case were such behaviour has been conjectured \cite{Coriano:2014gja} is in supersymmetric theories, where the superconformal anomaly manifests itself with the appearance of 3 anomaly poles. These affect vertices containing one superconformal and two (super)vector currents, which cover both the AVV and the TJJ cases, plus a third anomaly vertex with the insertion of a supersymmetric current. Also in this case it is suggestive to interpret such exchanges as due to interpolating effective axion/dilaton/axino interactions. Obviously, it remains an open issue whether such behaviour is an indication of the existence of a phase of the theory in which supersymmetry is nonlinearly realized. In such a case such composite intermediate states could become asymptotic, being the Goldstone modes of a broken superconformal symmetry.\\ 
It is quite interesting that recent analysis in solid state theory have suggested that such massless exchanges in the 
chiral and conformal anomaly actions play an important role in the theory of topological materials \cite{Rinkel:2016dxo,Chernodub:2017jcp}, confirming previous analysis in high energy 
\cite{Giannotti:2008cv,2009PhLB..682..322A,Armillis:2009pq,Armillis:2009sm}. Such universal behaviour is related to the fundamental role played by the anomalies in quantum field theory.
 
  \subsection{The perturbative $TJJ$ vertex} 
 The perturbative cases discussed in the past, concerning this vertex, cover QED, QCD and the neutral currents sector of the Standard Model \cite{Coriano:2011zk}, where the features described above are evident at one-loop. Even in the presence of a broken (massive) phase, in a mass-independent regularization scheme such as dimensional regularization, it is still possible to identify anomaly poles in this correlator, which are present in each gauge-invariant sector. \\
 In QCD, for instance, the two gauge invariant sectors involve at one-loop either quarks or gluons, and the pattern that we are going to describe is separately present in each of these two sectors. We refer to \cite{Coriano:2014gja} for a general and combined analysis of such features and to \cite{Coriano:2011ti} for a complete analysis of the neutral currents sector of the Standard Model.
 We briefly summarize the status of this analysis in the case of QED.
 
 The $TJJ$ vertex, in QED, describes the coupling of a graviton to two photons and is a source of the conformal anomaly. Perturbative investigations of this correlator have shown that the pole contribution is described, in the 1-particle irreducible effective action, by the term 
 \be
 \label{pole}
\mathcal{S}_{pole}= - \frac{e^2}{ 36 \pi^2}\int d^4 x d^4 y \left(\square h(x) - \partial_\mu\partial_\nu h^{\mu\nu}(x)\right)  \square^{-1}_{x\, y} F_{\alpha\beta}(x)F^{\alpha\beta}(y)
\ee

which can be extracted from the 1-loop expression of the vertex, using a suitable decomposition.  
In fact, the amplitude for the $TJJ$ can be expanded in the basis proposed by \cite{Giannotti:2008cv}, in terms of 13 independent tensors structures given in Table (\ref{genbasis}). It can be written as
\begin{equation}
\Gamma^{\m_1\n_1\m_2\m_3}(p_2,p_3)=\sum_{i=1}^{13}\,F_i(s;s_1,s_2,0)\,t_i^{\m_1\n_1\m_2\m_3}(p_2,p_3),
\end{equation} 
where the invariant amplitudes $F_i$ are functions of the kinematic invariants $s=p_1^2=(p_2+p_3)^2$, $s_1=p_2^2$, $s_2=p_3^2$, and the $t_i^{\m_1\n_1\m_2\m_3}$ define the basis of the independent tensor structures. 

\begin{figure}[t]
\begin{center}
\includegraphics[scale=0.9]{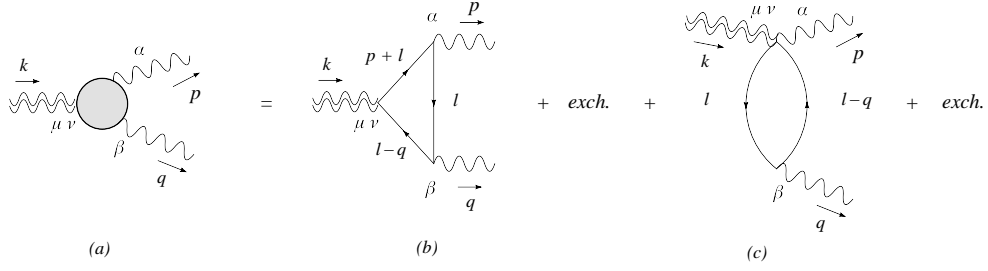}
\caption{\small The complete $TJJ$ one-loop vertex (a) given by the sum of the 1PI contributions with triangle (b) and pinched topologies (c). }
\label{vertex}
\end{center}
\end{figure}

\begin{table}
$$
\begin{array}{|c|c|}\hline
i & t_i^{\mu\nu\alpha\beta}(p,q)\\ \hline\hline
1 &
\left(k^2 g^{\mu\nu} - k^{\mu } k^{\nu}\right) u^{\alpha\beta}(p.q)\\ \hline
2 &
\left(k^2g^{\mu\nu} - k^{\mu} k^{\nu}\right) w^{\alpha\beta}(p.q)  \\ \hline
3 & \left(p^2 g^{\mu\nu} - 4 p^{\mu}  p^{\nu}\right)
u^{\alpha\beta}(p.q)\\ \hline
4 & \left(p^2 g^{\mu\nu} - 4 p^{\mu} p^{\nu}\right)
w^{\alpha\beta}(p.q)\\ \hline
5 & \left(q^2 g^{\mu\nu} - 4 q^{\mu} q^{\nu}\right)
u^{\alpha\beta}(p.q)\\ \hline
6 & \left(q^2 g^{\mu\nu} - 4 q^{\mu} q^{\nu}\right)
w^{\alpha\beta}(p.q) \\ \hline
7 & \left[p\cdot q\, g^{\mu\nu}
-2 (q^{\mu} p^{\nu} + p^{\mu} q^{\nu})\right] u^{\alpha\beta}(p.q) \\ \hline
8 & \left[p\cdot q\, g^{\mu\nu}
-2 (q^{\mu} p^{\nu} + p^{\mu} q^{\nu})\right] w^{\alpha\beta}(p.q)\\ \hline
9 & \left(p\cdot q \,p^{\alpha}  - p^2 q^{\alpha}\right)
\big[p^{\beta} \left(q^{\mu} p^{\nu} + p^{\mu} q^{\nu} \right) - p\cdot q\,
(g^{\beta\nu} p^{\mu} + g^{\beta\mu} p^{\nu})\big]  \\ \hline
10 & \big(p\cdot q \,q^{\beta} - q^2 p^{\beta}\big)\,
\big[q^{\alpha} \left(q^{\mu} p^{\nu} + p^{\mu} q^{\nu} \right) - p\cdot q\,
(g^{\alpha\nu} q^{\mu} + g^{\alpha\mu} q^{\nu})\big]  \\ \hline
11 & \left(p\cdot q \,p^{\alpha} - p^2 q^{\alpha}\right)
\big[2\, q^{\beta} q^{\mu} q^{\nu} - q^2 (g^{\beta\nu} q^ {\mu}
+ g^{\beta\mu} q^{\nu})\big]  \\ \hline
12 & \big(p\cdot q \,q^{\beta} - q^2 p^{\beta}\big)\,
\big[2 \, p^{\alpha} p^{\mu} p^{\nu} - p^2 (g^{\alpha\nu} p^ {\mu}
+ g^{\alpha\mu} p^{\nu})\big] \\ \hline
13 & \big(p^{\mu} q^{\nu} + p^{\nu} q^{\mu}\big)g^{\alpha\beta}
+ p\cdot q\, \big(g^{\alpha\nu} g^{\beta\mu}
+ g^{\alpha\mu} g^{\beta\nu}\big) - g^{\mu\nu} u^{\alpha\beta} \\
& -\big(g^{\beta\nu} p^{\mu}
+ g^{\beta\mu} p^{\nu}\big)q^{\alpha}
- \big (g^{\alpha\nu} q^{\mu}
+ g^{\alpha\mu} q^{\nu }\big)p^{\beta}  \\ \hline
\end{array}
$$
\caption{The basis of 13 fourth rank tensors satisfying the vector current conservation on the external lines with momenta $p$ and $q$. \label{genbasis}}
\end{table}
On this basis, which is built by imposing on the $TJJ$ vertex all the Ward identities derived from diffeomorphism invariance and gauge invariance, it is possible to use Bose symmetry and conservation WI's to reduce the number of form factors.  

\subsection{The structure of the (partially transverse) $F$-basis} 
The set of the $13$ tensors $t_i$ is linearly independent for generic $k^2, p^2, q^2$
different from zero. Five of the $13$ are Bose symmetric, 
\be
t_i^{\mu\nu\alpha\beta}(p,q) = t_i^{\mu\nu\beta\alpha}(q,p)\,,\qquad i=1,2,7,8,13\,,
\ee
while the remaining eight tensors are Bose symmetric pairwise
\bea
\label{pair}
&&t_3^{\mu\nu\alpha\beta}(p,q) = t_5^{\mu\nu\beta\alpha}(q,p)\,,\\
&&t_4^{\mu\nu\alpha\beta}(p,q) = t_6^{\mu\nu\beta\alpha}(q,p)\,,\\
&&t_9^{\mu\nu\alpha\beta}(p,q) = t_{10}^{\mu\nu\beta\alpha}(q,p)\,,\\
&&t_{11}^{\mu\nu\alpha\beta}(p,q) = t_{12}^{\mu\nu\beta\alpha}(q,p)\,.
\eea
In the set are present two tensor structures
\bes\bea
&&u^{\alpha\beta}(p,q) \equiv (p\cdot q) g^{\alpha\beta} - q^{\alpha}p^{\beta}\,,\\
&&w^{\alpha\beta}(p,q) \equiv p^2 q^2 g^{\alpha\beta} + (p\cdot q) p^{\alpha}q^{\beta}
- q^2 p^{\alpha}p^{\beta} - p^2 q^{\alpha}q^{\beta}\,,
\eea \label{uwdef}\ees
which appear in $t_1$ and $t_2$ respectively.
Each of them satisfies the Bose symmetry  requirement,
\bes\bea
&&u^{\alpha\beta}(p,q) = u^{\beta\alpha}(q,p)\,,\\
&&w^{\alpha\beta}(p,q) = w^{\beta\alpha}(q,p)\,,
\eea\ees
and vector current conservation,
\bes\bea
&&p_{\alpha} u^{\alpha\beta}(p,q) = 0 = q_{\beta}u^{\alpha\beta}(p,q)\,,\\
&&p_{\alpha} w^{\alpha\beta}(p,q) = 0 = q_{\beta}w^{\alpha\beta}(p,q)\,.
\eea\ees
They are obtained from the variation of gauge invariant quantities
$F_{\mu\nu}F^{\mu\nu}$ and $(\partial_{\mu} F^{\mu}_{\ \,\lambda})(\partial_{\nu}F^{\nu\lambda})$

\bea
&&u^{\alpha\beta}(p,q) = -\frac{1}{4}\int\,d^4x\,\int\,d^4y\ e^{ip\cdot x + i q\cdot y}\ 
\frac{\delta^2 \{F_{\mu\nu}F^{\mu\nu}(0)\}} {\delta A_{\alpha}(x) A_{\beta}(y)} \,,
\label{one}\\
&&w^{\alpha\beta}(p,q) = \frac{1}{2} \int\,d^4x\,\int\,d^4y\ e^{ip\cdot x + i q\cdot y}\
\frac{\delta^2 \{\partial_{\mu} F^{\mu}_{\ \,\lambda}\partial_{\nu}F^{\nu\lambda}(0)\}} 
{\delta A_{\alpha}(x) A_{\beta}(y)}\,.\label{two}
\eea\label{three}
 All the $t_i$'s are transverse in their photon indices
\bea
q^\alpha t_i^{\mu\nu\alpha\beta}=0  \qquad p^\beta t_i^{\mu\nu\alpha\beta}=0.
\eea
$t_2\ldots t_{13}$ are traceless, $t_1$ and $t_2$ have trace parts in $d=4$. With this decomposition, the two vector Ward identities are automatically satisfied by all the amplitudes, as well as the Bose symmetry. \\ 
Diffeomorphism invariance, instead, is automatically satisfied (separately) by the two tensor structures $t_1$ and $t_2$, which are completely transverse, while it has to be imposed on the second set ($t_3\ldots t_{13}$). Such identities are   
 
\bea
&& - p^2 F_3 + (3 q^2 + 4 p\cdot q) F_5 + (2 p^2 + p\cdot q) F_7 - p^2 q^2 F_{10}
- p^2 (p^2 + p\cdot q) F_9 + p^2 q^2 F_{11} = 0\,, \label{WIcons1a}\nn 
&& p^2 F_4 - (3 q^2 + 4 p\cdot q) F_6 - (2 p^2 + p\cdot q) F_8 - p\cdot q F_{10}
+ (q^2 + 2 p\cdot q) F_{11} = 0\,,\label{WIcons1b}\nn
&& -p\cdot q \,(p^2 + p\cdot q) F_9 - q^2 (q^2 + p\cdot q) F_{11} + F_{13} + \Pi(p^2)  =0 \,,
\label{cwi}\eea

with $\Pi(p^2)$ being the scalar 2-point function of momentum $p$, and a symmetric set of 3 equations obtained from (\ref{cwi}) by exchanging $p$ with $q$ and using the pair relations (\ref{pair}). In this way it is possible to extract from the 9 traceless tensor structures a (completely) transverse and traceless set of 5 amplitudes, which will be given below, two of them related by the bosonic symmetry. \\
To summarize, from the original 13 tensor structures $t_i$, split into a set of two transverse and trace components and a remaining set of 11 partially transverse but traceless ones (in $d=4$), one is left with 7 form factors after imposing the pairing conditions \eqref{pair}. Finally, imposing the conservations WI's \eqref{cwi} these are reduced to 4, which are related to the 4 form factors $A_i$'s  introduced by BMS in their reconstruction method \cite{Bzowski:2013sza}. \\
\subsection{The transverse traceless basis}
 It is possible to show that these amplitudes are in a one-to-one correspondence with the form factors $A_j$ $(j=1,\ldots 4)$  introduced in the parameterization of the $TJJ$ correlator presented in \cite{Bzowski:2013sza}. In that work the full 3-point function is parameterized in terms of transverse (with respect to all the external momenta) traceless components plus extra terms identified via longitudinal Ward identities of the $TJJ$ (the so-called {\em local} or {\em contact terms}) characterised by pinched topologies 
\begin{align}
\braket{\braket{T^{\mu_1\nu_1}\,J^{\mu_2}\,J^{\mu_3}}}&=\braket{\braket{t^{\mu_1\nu_1}\,j^{\mu_2}\,j^{\mu_3}}}+ \textrm{local terms}.\end{align}
Here we have switched to a symmetric notation for the external momenta, with $(p_1,p_2,p_3)\equiv (k,p,q)$, and 
with the transverse traceless parts expanded in terms of a set of the form factors $A_j$ mentioned above  
\begin{align}
\langle t^{\mu_1\nu_1}(p_1)j^{\mu_2}(p_2)j^{\mu_3}(p_3)\rangle& =
{\Pi_1}^{\mu_1\nu_1}_{\alpha_1\beta_1}{\pi_2}^{\mu_2}_{\alpha_2}{\pi_3}^{\mu_3}_{\alpha_3}
\left( A_1\ p_2^{\alpha_1}p_2^{\beta_1}p_3^{\alpha_2}p_1^{\alpha_3} + 
A_2\ \delta^{\alpha_2\alpha_3} p_2^{\alpha_1}p_2^{\beta_1} + 
A_3\ \delta^{\alpha_1\alpha_2}p_2^{\beta_1}p_1^{\alpha_3}\right. \notag\\
& \left. + 
A_3(p_2\leftrightarrow p_3)\delta^{\alpha_1\alpha_3}p_2^{\beta_1}p_3^{\alpha_2}
+ A_4\  \delta^{\alpha_1\alpha_3}\delta^{\alpha_2\beta_1}\right).\label{DecompTJJ}
\end{align}
In this expression ${\Pi_1}^{\mu_1\nu_1}_{\alpha_1\beta_1}$ is a transverse and traceless projector built out of momentum $p_1$, while  ${\pi_2}^{\mu_2}_{\alpha_2}$ and ${\pi_3}^{\mu_3}_{\alpha_3}$ denote transverse projectors respect to the momenta $p_2$ and $p_3$.  \\ 
Coming to the explicit form of the $A_j$, these can be determined, modulo some constants, by the solution of the primary Ward identities. Primary Ward identities are second order (vector) differential constraints on a tensor correlator which are reformulated as a set of scalar equations \cite{Bzowski:2013sza}. They are obtained  by the action of the generators of the special conformal transformations $(K^\kappa)$ on the $TJJ$ amplitude. \\
 In momentum space, after an involved analysis, one obtains a set of scalar equations for the $A_j$, whose primary Ward identities are formulated in terms of a set of second order scalar operators

\begin{align}
K_i&=\sdfrac{\partial^2}{\partial p_i^2}+\sdfrac{d+1-2\Delta_i}{p_i}\,\sdfrac{\partial}{\partial p_i},\quad i=1,2,3\\
K_{ij}&=K_i-K_j,
\end{align}
where $\D_i$ is the conformal dimension of the i-th operator in the 3-point function under consideration. In our case, for the $\braket{TJJ}$,  $\Delta_1=d$, $\D_2=\Delta_3=d-1$, and the primary CWI's take the form
\begin{equation}
\begin{split}
0&=K_{13}A_1\\
0&=K_{13}A_2+2A_1\\
0&=K_{13}A_3-4A_1\\
0&=K_{13}A_3(p_2\leftrightarrow p_3)\\
0&=K_{13}A_4-2A_3(p_2\leftrightarrow p_3)
\end{split}
\hspace{1.5cm}
\begin{split}
0&=K_{23}A_1\\
0&=K_{23}A_2\\
0&=K_{23}A_3-4A_1\\
0&=K_{23}A_3(p_2\leftrightarrow p_3)+4A_1\\
0&=K_{23}A_4+2A_3-2A_3(p_2\leftrightarrow p_3).
\end{split}\label{Primary}
\end{equation}
The solutions of such equations are expressed in terms of linear combinations of generalized hypergeometric functions of two variables (Appel's functions $F_4$), recently solved in terms of parametric integrals of three Bessel functions (3-K integrals), 
as discussed in \cite{Bzowski:2013sza,Bzowski:2015yxv}. A direct analysis of the solutions using the Fuchsian properties 
of these equations will be presented elsewhere \cite{inpreparation1}.\\
The tensor nature of the correlator implies that also some first order differential constraints need to be imposed (called secondary CWI's in \cite{Bzowski:2013sza}). The solution of such constraints, however, can be performed at special kinematic points (for instance at equal invariant mass of the two photons, $p_2^2=p_3^2$, or, alternatively, in the massless limit of the graviton line), which constrain the undetermined constants of the general solutions of the primary CWI's (\ref{Primary}). Secondary CWI's are related to longitudinal/trace terms of the correlators, and henceforth to contact terms.\\

\section{The renormalization of $F_{13}$: $d$-dimensional analysis }
In order to clarify the connection between the apperance of a pole and the process of renormalization, we consider the $d-$dimensional structure of the CWI's of the $TJJ$ in the $F-$basis. QED provides a realization of this behaviour at one-loop and we will stick to this example for definiteness, in order to clarify our discussion.

The $TJJ$ correlator in QED is conformal in $d$ dimension, with finite form factors which are dimensionally regulated and therefore it does not develop any conformal anomaly. We can use the $F$-basis to parameterize the correlator, now in $d$ dimensions, in terms of the same 13 form factors $F_i$ introduced before and of the corresponding tensor structures $t_i$.\\
 Notice that the separation of these 13 structures into trace-free and trace parts is valid only in $d=4$ for most of the structures, except for $t_9,t_{10},t_{11}$ and $t_{12}$, which remain traceless in $d$ dimensions. We are assuming that the contractions with the metric tensor is performed in $d$ dimensions with a metric 
$g_{\mu\nu}(d)$. The 4-dimensional metric, instead, will be denoted as $g_{\mu\nu}(4)$.\\
For instance, a contraction of $t_1$ and $t_2$ in d- dimensions will give 
\bea
g^{\mu\nu}(d)t_1^{\mu\nu\alpha\beta}&=&(d-4) k^2 u^{\alpha\beta}(p,q)\nn 
g^{\mu\nu}(d)t_2^{\mu\nu\alpha\beta}&=&(d-4) k^2 w^{\alpha\beta}(p,q),
\eea
and similarly for all the other structures, except for those mentioned above, which are trace-free in any dimensions.\\
 Using the completeness of the $F$-basis and by a direct analysis of the CWI's which will be detailed elsewhere, we can identify the mapping between the form factors of such basis and those of the $A$-basis. They are conveniently expressed in terms of the momenta $(p_1,p_2,p_3)$ in the form
 \bea
A_1&=&4(F_7-F_3-F_5)-2p_2^2F_9-2p_3^2F_{10}\nn
A_2&=&2(p_1^2-p_2^2-p_3^2)(F_7-F_5-F_3)-4p_2^2p_3^2(F_6-F_8+F_4)-2F_{13}\nn
A_3&=&p_3^2(p_1^2-p_2^2-p_3^2)F_{10}-2p_2^2\,p_3^2 F_{12}-2F_{13}\nn
A_3(p_2\leftrightarrow p_3)&=&p_2^2(p_1^2-p_2^2-p_3^2)F_9-2p_2^2p_3^2F_{11}-2F_{13}\nn
A_4&=&(p_1^2-p_2^2-p_3^2)F_{13},
\label{mapping1}
\eea
which are transverse and traceless, with $A_1$, $A_2$ and $A_4$ symmetric.  \\
Given the correspondence \eqref{mapping1}, it is worth noticing that the form factor $A_3$ and its corresponding symmetric $A_3(p_2\leftrightarrow p_3)$  are consistently defined in terms of the $F_i$, since in the F-set
\begin{equation}
F_{10}(s;s_1,s_2,0)=F_9(s;s_2,s_1,0),\quad F_{12}(s;s_1,s_2,0)=F_{11}(s;s_2,s_1,0),
\end{equation}
which shows the consistency of the mapping between the $F$ and $A$ basis. 

The presence of two tensor structures of nonzero trace in $d=4$ in the $F-$basis, however, is at first sight
slightly puzzling, since the correspondence between the appearance of an anomaly pole (and henceforth of a trace) and the process of renormalization does not seem to be unique. We are looking for a single (anomaly) pole whose origin should be traced back to renormalization. The expansion may allow extra poles, but they will be unrelated to renormalization.
We are going to show that indeed there are no extra poles sharing such a feature.\\
The two sets $A_j$ and $F_i$  differ in several ways, and emphasize different properties of the same $TJJ$ correlator. On the one hand, the $F$-basis sheds light on the 
origin of the anomaly pole, as we are going to show below, by linking its origin to the single form factor $F_{13}$  which exhibits a divergence and requires renormalization. \\
Notice that the result for the $A_i$'s presented in \cite{Bzowski:2017poo} shows, by an analysis of the 
3-K integrals, that the singularities of the $A_i$'s are those of $A_2,A_3$ and $A_4$. This is consistent 
with the mapping \eqref{mapping1} since those are exactly the combinations in which the divergent form factor $F_{13}$ appears. \\
This specific origin of the singularity, which can be directly identified in the F-basis, is not directly manifest in the $A-$basis. The $A_i$' s, on the other hand, describe a minimal set of form factors which are suitable for resolving the CWI's of the correlators, but shadow the origin of the singular 
behaviour, since 3 out of 4 of them manifest UV singularities and need to be renormalized. By using the $F-$basis, instead we know where to look for singularities in a rather simple way, this is $F_{13}$.

\subsection{The anomaly pole from renormalization}
In order to trace back the origin of the anomaly pole in $TJJ$, starting from $d$-dimension and using the $F$-basis, we request that this corelator has no trace (i.e. be anomaly free). The anomaly will emerge in dimensional regularization as we take the $d\to 4$ limit. The trace WI's provide the two key conditions that we need. In fact we obtain

\be
\label{one}
F_1=\frac{(d-4)}{p_1^2(d-1)}\big[F_{13}-p_2^2\,F_3-p_3^2\,F_5-p_2\cdot p_3\, F_7\big]\\
\ee
and 
\be
\label{two}
F_2=\sdfrac{(d-4)}{p_1^2(d-1)}\big[p_2^2\,F_4+p_3^2\,F_6+p_2\cdot p_3\,F_8\big].
\ee
Both equations are crucial in order to understand the way the renormalization procedure works for such correlator.
From Eq. (\ref{two}) it is clear that by sending $d\to 4$, $F_2$ vanishes,
\be
F_2=\frac{\epsilon}{(d-1) p_1^2}\big[p_2^2\,F_4+p_3^2\,F_6+p_2\cdot p_3\,F_8\big]\to 0,
\ee
for all the form factors $F_4,F_6$ and $F_8$ are finite for dimensional reasons. In fact, from the scaling dimensions of the corresponding tensor structures $t_4,t_6$ and $t_8$ one concludes that they are finite, and therefore $F_2$ is indeed zero in this limit, since the right hand side of \eqref{two} has no poles in $\epsilon\equiv d-4$. \\
At this stage, after the limiting procedure, at $d=4$ we are left in the $F-$basis with 4 independent combinations of form factors from the original 7 
(those given in \eqref{mapping1}), which are sufficient to describe the (complete) transverse traceless sector of the theory, plus an additional form factor $F_1$. Therefore, by taking the $d\to 4$ limit, the $F-$set contains only one single tensor structure (and an associated form factor) with a nonzero trace, which should account for the anomaly in $d=4$. This result is obviously confirmed in perturbation theory in QED \cite{Armillis:2009pq}.\\
As already mentioned, $F_{13}$ is the only form factor that needs to be renormalized in the $F$-set and it is characterized by the appearance of a single pole in $1/\epsilon$ in dimensional regularization. The fact that such singularity will be at all orders of the form $1/\epsilon$ and not higher is a crucial ingredient in the entire construction. Such assumption is expected to be consistent with the analysis in conformal field theory since the only available counterterm to regulate the theory is given by 
\be
\frac{1}{\epsilon}\int d^4 x \sqrt{g} F_{\mu\nu}F^{\mu\nu}
\ee
which renormalizes the 2-point function $\langle JJ \rangle$ and henceforth $F_{13}$. Explicit computations in QED show that 
\be
\label{f13}
F_{13}= G_0(p_1^2, p_2^2,p_3^3) -\frac{1}{2} \, [\Pi (p_2^2) +\Pi (p_3^2)]
\ee
with $G_0$ a lengthy expression which remains finite as $d\to 4$, with the origin of the singularity traced back to the 
scalar form factor $\Pi(p^2)$ of the photon 2-point function. For this purpose, we just recall that the structure of the two-point function of two conserved vector currents of scaling dimensions $\eta_1$ and $\eta_2$ is given by \cite{Coriano:2013jba}
\be
\label{TwoPointVector}
G_V^{\alpha \beta}(p) = \delta_{\eta_1 \eta_2}  \, c_{V 12}\, 
\frac{\pi^{d/2}}{4^{\eta_1 - d/2}} \frac{\Gamma(d/2 - \eta_1)}{\Gamma(\eta_1)}\,
\left( \eta^{\alpha \beta} -\frac{p^\alpha p^\beta}{p^2} \right)\
(p^2)^{\eta_1-d/2} \,,
\ee
with $c_{V12}$ being an arbitrary constant. It requires the two currents to share the same dimensions and manifests only a single pole in ${1/\epsilon}$.   
In dimensional regularization, in fact,  the divergence can be regulated with $d \to d - 2 \epsilon$. Expanding the product $\Gamma(d/2-\eta)\,(p^2)^{\eta - d/2}$, which appears in the two-point function, in a Laurent series around $d/2 - \eta = -n$ (integer) gives the single pole in $1/\epsilon$ behaviour  \cite{Coriano:2013jba}
\bea
\label{expansion}
\Gamma\left(d/2-\eta\right)\,(p^2)^{\eta-d/2} = \frac{(-1)^n}{n!} \left( - \frac{1}{\epsilon} + \psi(n+1)  + O(\epsilon) \right) (p^2)^{n + \epsilon} \,,
\eea
where $\psi(z)$ is the logarithmic derivative of the Gamma function, and $\epsilon$ takes into account the divergence of the two-point correlator for particular values of the scale dimension $\eta$ and of the space-time dimension $d$. Therefore, the divergence in $F_{13}$ is then given by a single pole in $\epsilon$ is of the form

\be 
\label{renf}
F_{13}=\frac{1}{d-4} \bar{F}_{13} + F_{13\, f}
\ee
In QED, for instance, one finds by an explicit computation that $\bar{F}_{13}=-e^2/(6 \pi^2)$ at one-loop and $\bar{F}_{13\, f}$ is finite \cite{Armillis:2009pq} and gets renormalized into $F_{13 R}$ only in its photon self-energy contributions \cite{Armillis:2009pq} $(s=p_1^2,\,s_1=p_2^2,\,s_2=p_3^2)$
\beqa
  { {F_{13,R} (s;\,s_1,\,s_2,\,0)}} &=&
  -\frac{1}{2} \left[ \Pi_R(s_1,0)+ \Pi_R(s_2,0) \right]
 + G_0(s,s_1,s_2) 
\eeqa
with 
\bea
 \Pi_R (s,0) = - \frac{e^2}{12 \,  \pi^2 } \, \left[\frac{5}{3} - \log \left(-\frac{s}{\mu^2} \right)\right],
 \eea
 denoting the renormalized scalar form factor of the $JJ$ correlator at one-loop and with $G_0$ defined in \eqref{f13}.

Inserting \eqref{renf} into (\ref{one}) we obtain
\be
\label{oneprime}
F_1=\frac{(d-4)}{p_1^2(d-1)}\left(  \frac{1}{d-4} \bar{F}_{13} + F_{13\, f}  -p_2^2\,F_3-p_3^2\,F_5-p_2\cdot p_3\, F_7\right),
\ee  
which in the $d\to 4$ limit gives, in general 
\be
F_1=\frac{\bar{F}_{13}}{3 p_1^2}
\ee
and specifically, in QED
\be
\label{stable}
F_1=- \frac{e^2}{32 \pi^2 s},
\ee
$(s\equiv k^2)$ showing that the anomaly pole in $F_1$ is indeed generated by the renormalization of the single divergent form factor 
$F_{13}$. In the case of QED, the relation between the prefactor in front of the $1/s$ pole and its relation to the QED $\beta$-function has been extensively discussed in \cite{Giannotti:2008cv,Armillis:2009pq}, to which we refer for further details. In performing the limit we have used the finiteness of the remaining form factors. 

\section{Implications in the non-abelian case} 
Further comparisons with the non-perturbative solutions of the CWI's, in the approach presented in \cite{Bzowski:2017poo} can be made using the perturbative results of \cite{Armillis:2010qk} for QCD, the pattern described above being still valid also in this case, 
although only the structure of the on-shell vertex, with $s$ arbitrary, but with the two gluons on shell ($s_1=s_2=0$) is available 
for a direct comparison. In the QCD case, as already mentioned, there are two anomaly poles, one for each gauge invariant sector.
While for the fermion loop the anomaly pole generated follows exactly the same pattern discussed above with minimal changes (modulo extra color factors), for the gluon loops another pole is present and shares quite similar 
features. In fact, in the gluon sector only one form factor gets renormalized, by choosing an appropriate basis, \cite{Armillis:2010qk} and the pattern that emerges in (\ref{oneprime}) is similar, with the obvious changes. Notice that in the non-abelian case the number of form factors in the transverse traceless sector of the correlator is still 4, and their expressions gets modified just by simple colour factors, with 4 of them affected by a single polar divergence in $1/(d-4)$, as pointed out in 
\cite{Bzowski:2017poo}. One can show that such divergences are again associated to the renormalization of the gluon 2-point function, appearing in a single form factor ($\phi_3$, in the notations of \cite{Armillis:2010qk}). 
 \section{Conclusions } 
We have proven that conformal anomaly poles are not the result of specific parameterizations of anomaly vertices, but are the natural signature of the anomaly, being related to renormalization. The solution of CWI's in momentum space, presented in recent analysis, are completely consistent with previous studies in abelian and non-abelian theories \cite{Giannotti:2008cv} \cite{Armillis:2009pq} \cite{Armillis:2010qk}. The phenomenon is therefore generic, and it is surely not limited to the high energy domain, but wherever anomaly actions are at work. Recent studies have underlined the important role played by such massless exchanges \cite{Rinkel:2016dxo,Chernodub:2017jcp} and of the anomaly 
in general \cite{Landsteiner:2013sja,Chernodub:2013kya}, in the context of the transport properties of topological insulators and of Weyl semimetals. In particular \cite{Rinkel:2016dxo,Chernodub:2017jcp} suggest physical realizations of the observations of \cite{Giannotti:2008cv} \cite{Armillis:2009pq,Armillis:2009im} concerning the structure of the chiral and conformal anomaly actions. 

A more detailed technical discussion of the results of this work will be presented by us elsewhere.  

\vspace{1.cm}

\centerline{\bf Acknowledgements} 
We thank Emil Mottola, Paul McFadden, Luigi Delle Rose and Kostas Skenderis for discussions. C.C. thanks Fiorenzo Bastianelli and Olindo Corradini for discussions and hospitality at the 
Universities of Bologna and Modena and Maxim Chernodub at the University of Tours (LMTP) for hospitality and discussions. Finally, he thanks the High Energy Theory group at ETH-Zurich for hospitality. 
This work is performed as part of the HEP-QFT research activity of INFN.



\end{document}